\documentclass[11pt]{article}

\usepackage[final]{acl}

\usepackage{times}
\usepackage{latexsym}

\usepackage[T1]{fontenc}

\usepackage[utf8]{inputenc}

\usepackage{microtype}

\usepackage{inconsolata}

\usepackage{graphicx}

\usepackage{cite}
\usepackage{amsmath,amssymb,amsfonts}
\usepackage{algorithmic}
\usepackage{textcomp}
\usepackage{xcolor}
\usepackage[]{hyperref}
\usepackage{subcaption}
\usepackage{makecell}

\title{GovScape: A Public Multimodal Search System for 70 Million Pages of Government PDFs}

\author{
\textbf{Ying-Hsiang Huang\textsuperscript{1}},
\textbf{Claire Gong\textsuperscript{1}},
\textbf{Shreya Shaji\textsuperscript{1}},
\textbf{Alison Yan\textsuperscript{1}},
\textbf{Leslie Harka\textsuperscript{1}},\\
\textbf{Albert Du\textsuperscript{1}}, 
\textbf{Anjali Gopal\textsuperscript{1}},
\textbf{Samuel J. Klein\textsuperscript{2}},
\textbf{Shannon Zejiang Shen\textsuperscript{3}},
\textbf{Mark Phillips\textsuperscript{4}},\\
\textbf{Trevor Owens\textsuperscript{5}},
\textbf{Kyle Deeds\textsuperscript{6}},
\textbf{Benjamin Charles Germain Lee\textsuperscript{1}}
\\
\\
 \textsuperscript{1}University of Washington,
 \textsuperscript{2}Harvard University,
\textsuperscript{3}Massachusetts Institute of Technology,\\
 \textsuperscript{4}University of North Texas,
 \textsuperscript{5}American Institute of Physics,
 \textsuperscript{6}Boston University
\\
 \small{
   \textbf{Correspondence:} \href{mailto:email@domain}{bcgl@uw.edu}
 }
}

\begin{document}
\maketitle
\begin{abstract}
Efforts over the past three decades have produced web archives containing billions of webpage snapshots and petabytes of data. The End of Term Web Archive alone contains millions of PDFs produced by the federal government. While preservation with web archives has been successful, significant challenges for access and discoverability remain. In this paper, we introduce GovScape, a public search system that supports multimodal searches across 10,015,993 federal government PDFs from the 2020 End of Term crawl (70,958,487 total PDF pages) -- to our knowledge, all renderable PDFs in the 2020 crawl that are 50 pages or under. GovScape supports four primary forms of search: in addition to providing (1) filter conditions over metadata facets including domain and crawl date and (2) exact text search against the PDF text, we provide (3) semantic text search and (4) visual search against the PDFs across individual pages, enabling users to structure queries such as ``redacted documents'' or ``pie charts.'' We detail GovScape's search affordances, embedding pipeline, system architecture, and open source codebase. 
Significantly, the total estimated compute cost for GovScape's pre-processing pipeline for 10 million PDFs was approximately \$1,500, equivalent to 47,000 PDF pages per dollar spent on compute, demonstrating the potential for immediate scalability. We evaluate GovScape by (1) analyzing 1,679 search queries and (2) benchmarking vector and keyword index efficiency using these queries. GovScape can be found at \textbf{\textcolor{blue}{\url{https://www.govscape.net}}.}
\end{abstract}

\section{Introduction}

Longstanding efforts by cultural heritage institutions such as the Internet Archive and the Library of Congress have yielded enormously rich web archives over the past two decades \citep{milligan_book}. These web archives represent an unparalleled opportunity to study the history of the 21st century. The End of Term Web Archive is unique among them due to its size as a public domain web archive and its historical significance, consisting of expansive crawls of federal government websites at the end of every presidential administration since 2004 \citep{phillips_2023_JCDL}. It is a rich source for journalists, academic researchers, and the public. 

\begin{figure*}[t]
    \centering
    \includegraphics[width=0.8\linewidth]{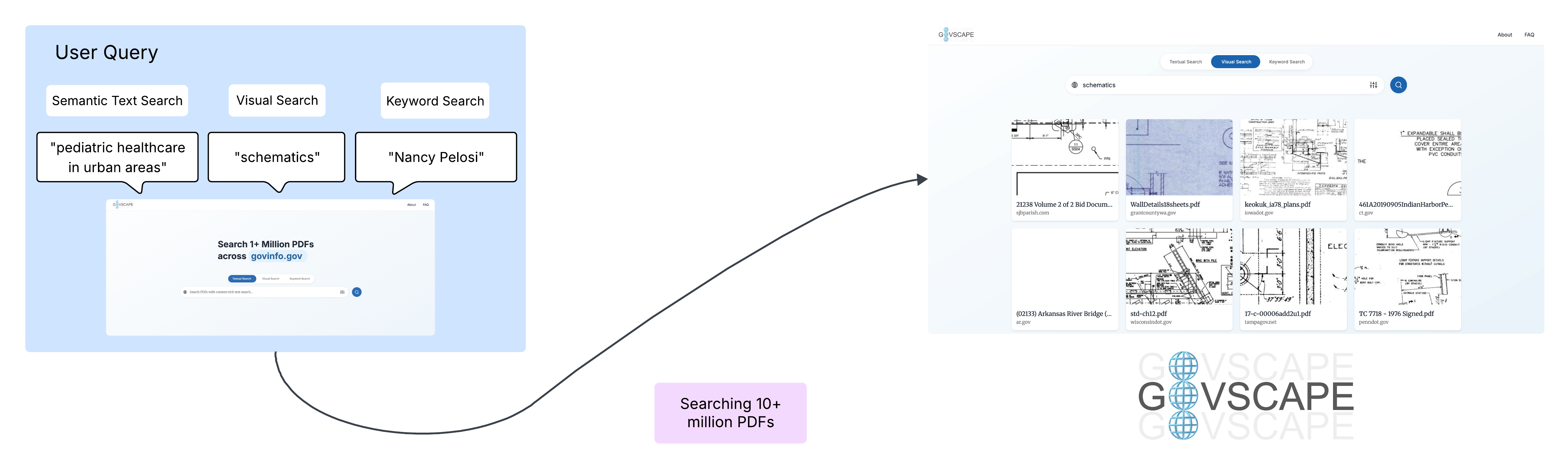}
    \caption{Our public search system supports three types of search over 10,015,993 million government PDFs (70,958,487 PDF pages): 1) semantic text search over PDF text, 2) visual search over individual PDF pages (treated as images), and 3) keyword search over PDF text, all of which can be applied with filter conditions against metadata.}
    \label{fig:overview}
\end{figure*}

Despite the availability of bulk data, the End of Term Web Archive – like all web archives of its size – remains difficult to search \citep{go_fish}. Currently, to search the PDFs, users must navigate the basic metadata facets (i.e., domain and crawl date) by downloading and querying massive CDX files and then downloading individual PDFs, or performing basic keyword search through the Internet Archive’s search functionality \citep{internet_archive_search}. End-users face this challenge with web archives writ large. Given the centrality of these archives to understanding the federal government in the 21st century, the importance of addressing this challenge at scale is redoubled. 

Recent research within library and information science as well as the digital humanities has demonstrated the value of multimodal models such as CLIP \citep{CLIP_paper} for searching digital cultural heritage collections \citep{mahowald_lee, smits_JOHD, smits_wevers, smits_kestemont, wevers_photos} (for a more detailed list of related work, we refer the reader to Appendix \ref{sec:related-work}). Previous work has also documented that born-digital government PDFs are not just textual but visual documents \citep{lee_owens}. For example, these PDFs contain redactions, figures, presentation slides, and form structures, all of which are visual in nature. Moreover, allowing users to formulate semantic natural language queries beyond keyword search (in which documents containing the exact search string or similar ones are returned) enables wider searches with higher precision. End-users should be able to formulate expressive multimodal queries across the PDFs in the End of Term Web Archive in particular and files in web archives more generally.

We present GovScape, a publicly-deployed search system containing 10 million PDFs from the 2020 End of Term crawl, covering the online presence of the federal government under the first Trump administration. In Figure \ref{fig:overview}, we show an overview of GovScape, including example queries that can be performed.
Built with a scalable back-end consisting of Faiss \citep{faiss},
our implementation of GovScape supports multiple types of searches over the textual and visual features of these PDFs: semantic text search, visual search over PDF pages (treated as images), keyword search (i.e., exact text search). All of these search methods can be combined with metadata filtering over facets from the CDX files. A demo video can be found here: \url{https://youtu.be/VpyiYW0nWp4}.

The contributions of our paper are as follows:
\begin{enumerate}
\itemsep0em 

    \item We introduce GovScape, a publicly-deployed, multimodal search system for 10 million government PDFs (70 million pages) from the 2020 crawl in the End of Term Web Archive. GovScape is available at:\\
    \textcolor{blue}{\url{https://www.govscape.net}}.
    \item We detail our multimodal  affordances, including visual search and semantic text search.
    \item We describe our development and public deployment of GovScape, including our scalable embedding pipeline and system architecture. Our estimated compute cost for processing all 10 million PDFs is approximately \$1,500.
    \item We release all of our open-source code at: {\color{blue}{\url{https://github.com/govscape/govscape/}}}. This code is extensible to other information retrieval tasks and datasets.
    \item We perform two forms of evaluation: log analysis of 1,679 end-user search queries, and vector \& keyword index benchmarking using these queries.
\end{enumerate}

\begin{figure*}[t]
    \centering    
    \includegraphics[width=0.7\linewidth]{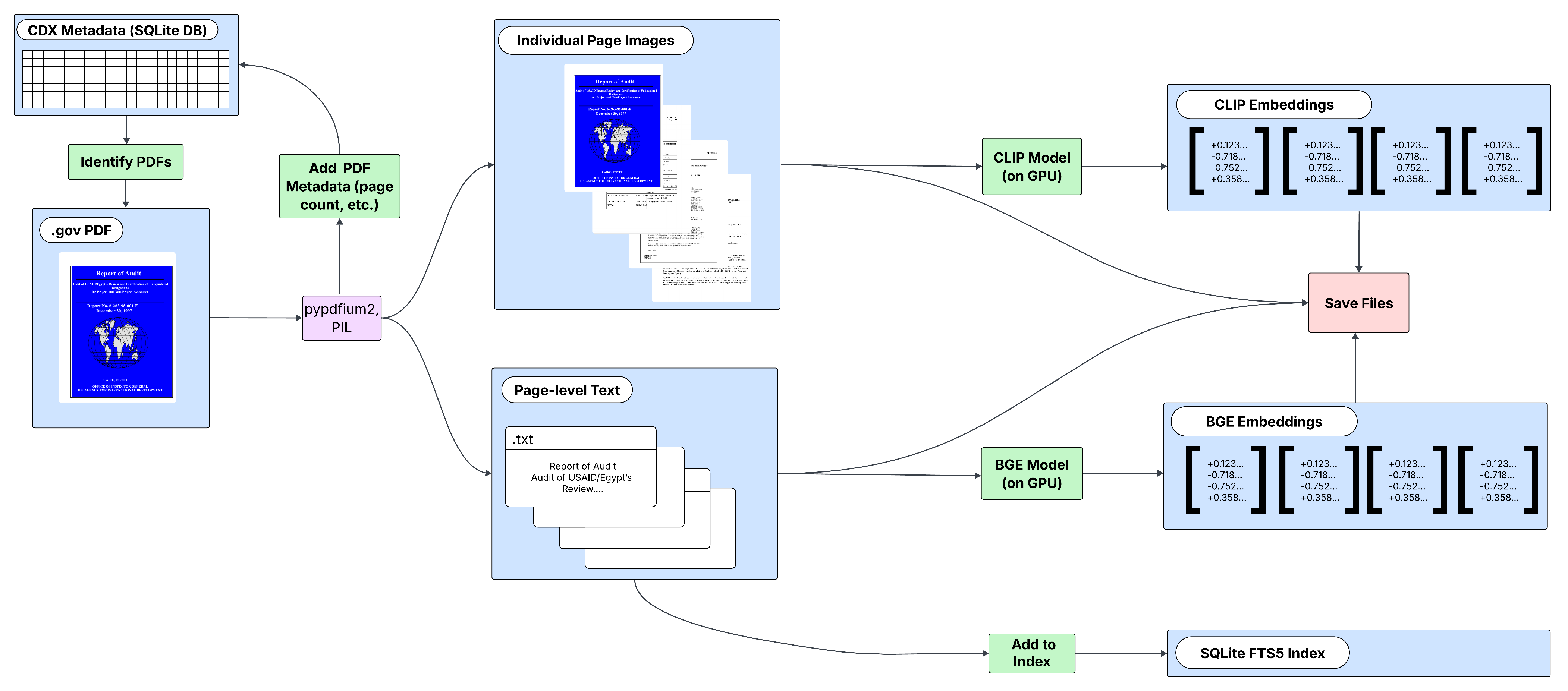}
    \caption{An overview of the GovScape pre-processing pipeline, showing how a single PDF in GovScape is parsed and semantified.}
    \label{fig:pipeline}
\end{figure*}

\section{Pre-Processing Pipeline}

\subsection{PDFs Included in GovScape}

We isolated PDFs within the End of Term Web Archive's 2020 crawl by identifying files listed within the CDX indices that have a file extension of .pdf or a MIME type of \texttt{application/pdf}. We deduplicated PDFs by hash, resulting in 10,532,521 total renderable PDFs. We then filtered out 516,528 PDFs with a page count above 50 pages to improve efficiency in our pipeline. The resulting 10,015,993 PDFs from 24,877 different domains were sent to our pipeline and included in GovScape. In total, this amounts to 70,958,487
pages of government PDFs. In Table \ref{tab:agency}, we break down the PDFs included within GovScape by domain.

\begin{table}
  \small
  \centering
\begin{tabular}{|c|c|c|}
\hline
\textbf{Domain} & \textbf{\# of PDF pages} & \textbf{\# of PDFs} \\
\hline \hline
sec.gov & 7,324,033 & 791,197 \\
\hline
govinfo.gov	& 3,051,220 & 1,078,761 \\
\hline
ca.gov & 2,860,263 & 375,405 \\
\hline
bls.gov & 2,261,400 & 234,667 \\
\hline 
wa.gov & 2,210,997 & 248,777 \\
\hline 
usda.gov & 2,149,605 & 311,660 \\
\hline 
epa.gov & 1,990,902 & 186,806 \\
\hline 
ed.gov  & 1,978,199 & 157,778 \\
\hline 
nasa.gov & 1,954,285 & 192,714 \\
\hline
noaa.gov & 1,770,684 & 195,122 \\
\hline
\end{tabular}
\caption{A breakdown of the 10 most prevalent domains in GovScape. Counts in this table (and this table alone) reflect PDFs before de-duplication.}\label{tab:agency}
\end{table}

\subsection{Pre-processing Pipeline}\label{sec:preprocessing}

Our pre-processing pipeline for GovScape is shown in Figure \ref{fig:pipeline}. It consists of the following steps:
\begin{enumerate}
\itemsep0em 
\item \textbf{PDF Identification.} We identify PDFs using the CDX indices for the 2020 End of Term crawl, which we migrated into a SQLite database. We then extract the PDFs from the corresponding WARC files.
\item \textbf{PDF Parsing.} We split each PDF into individual pages, generating page-level images and extracting the page-level text embedded within the PDF file using \texttt{pypdfium2} and \texttt{PIL}. We exclude pages with no text embedded within the PDF (it might only have images or be a digitized document without OCR, etc.).
\item \textbf{Text Embedding.} We embed the text for each page using the \texttt{BAAI/bge-base-en-v1.5} model \citep{bge_embedding} chosen for its balance of cost and performance as demonstrated in the MTEB leaderboard \citep{mteb1, mteb2}. This model has a context window of 512 tokens, and we pass the first 512 tokens from the PDF page.
\item \textbf{CLIP Embedding.} We embed each page-level image using the embedding model \texttt{openai/clip-vit-base-patch32}  \citep{CLIP_paper} for visual recognition. 
\item \textbf{Metadata Generation.} We extract relevant metadata for the PDF from the End of Term CDX files. This includes metadata fields such as URL and crawl date. We further enrich this with the number of pages in the PDF.
\item \textbf{Keyword Indexing.} We add the page-level full text to our Lucene index, which supports traditional keyword searches over text.
\end{enumerate}

\begin{figure*}[t!]
    \centering
    \begin{subfigure}[t]{0.45\textwidth}
    \centering
    \includegraphics[width=\linewidth]{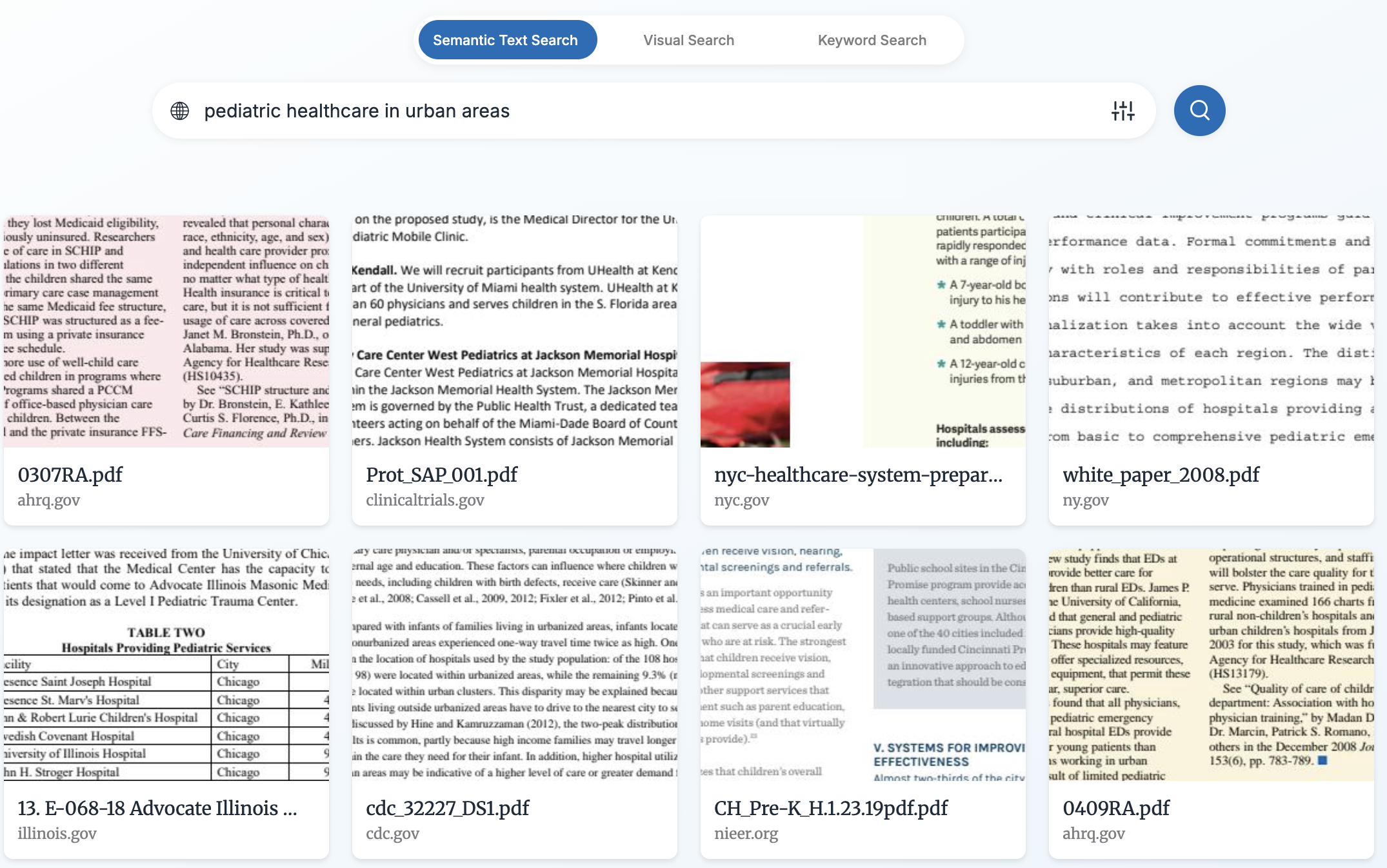}
    \caption{A semantic text search: ``pediatric healthcare in urban areas.'' 
    }
    \label{fig:semantic_text_example}
    \end{subfigure}
    \begin{subfigure}[t]{0.45\textwidth}
    \centering
    \includegraphics[width=\linewidth]{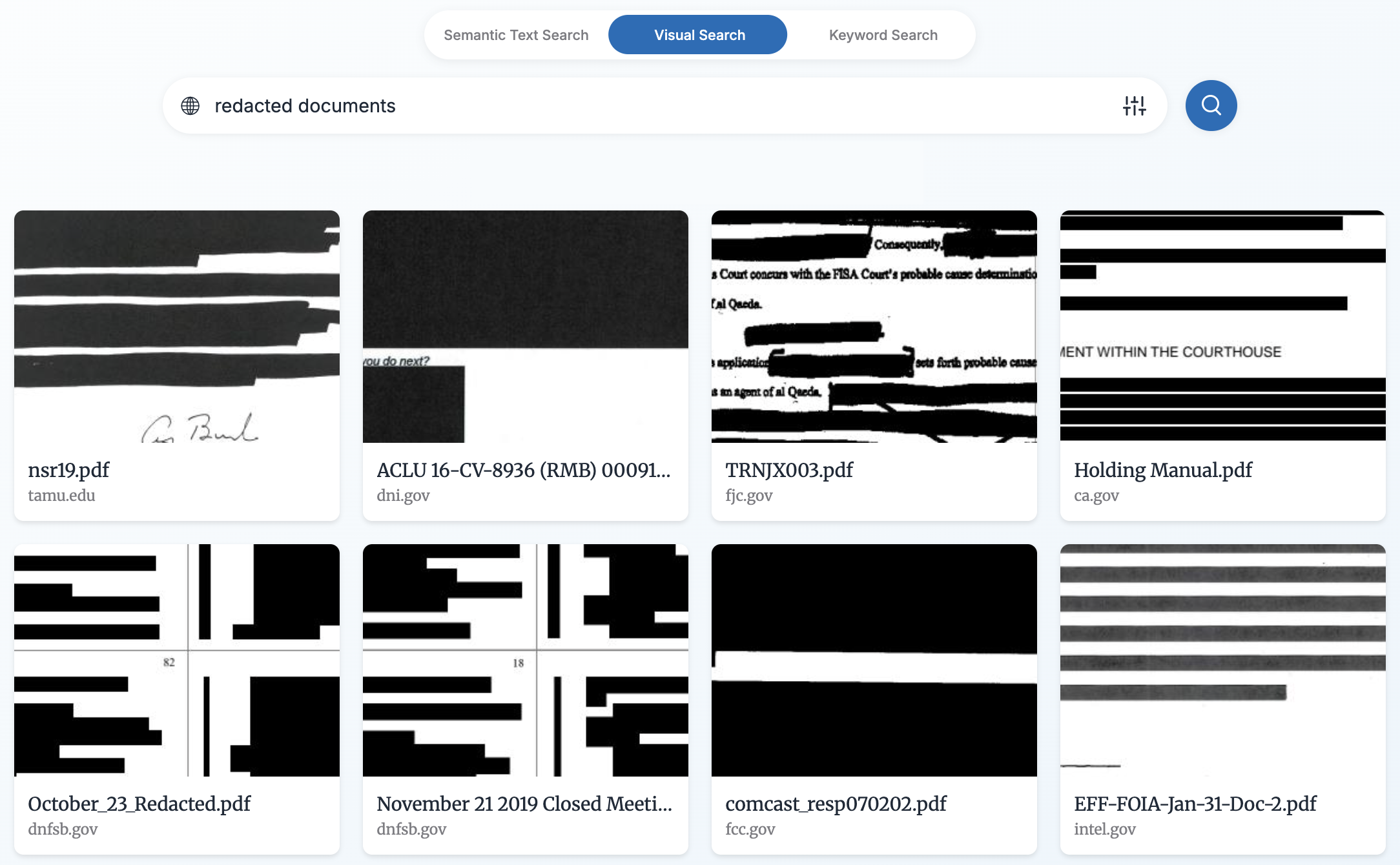}
    \caption{A visual search: ``redacted documents."}
    \label{fig:visual_example}
    \end{subfigure}
    \caption{Examples of semantic text search (Figure \ref{fig:semantic_text_example}) and visual search (Figure \ref{fig:visual_example}) in GovScape.}\label{fig:example_searches}
\end{figure*}

\begin{table}[htp]
\small
\centering
\begin{tabular}{|c|c|c|c|}
\hline
\textbf{Task Type} & \textbf{Cost} \\
\hline
\hline
\makecell{Pipeline Minus Indexing \\ (prior run w/ 4,736,080 PDFs)}& \$620.87 \\
\hline
\multicolumn{1}{|l|}{\hspace{5mm} 1. List} &  \\
\hline
\multicolumn{1}{|l|}{\hspace{5mm} 2. Download} &  \\
\hline
\multicolumn{1}{|l|}{\hspace{5mm} 3. PDF to Text \& Image} &  \\
\hline
\multicolumn{1}{|l|}{\hspace{5mm} 4. Text Embedding} &  \\
\hline
\multicolumn{1}{|l|}{\hspace{5mm} 5. Image Embedding} &  \\
\hline
\multicolumn{1}{|l|}{\hspace{5mm} 6. Metadata} &  \\
\hline
\multicolumn{1}{|l|}{\hspace{5mm} 7. Upload} &  \\
\hline
\makecell{\textbf{Pipeline Minus Indexing} \\ \textbf{(Extrapolated)}} & \textbf{\$1,313.02}\\
\hline 
\hline
\makecell{\textbf{FAISS Index Building (all PDFs)}} & \textbf{\$65.79} \\
\hline 
\hline 
\makecell{\textbf{Keyword Index Building (all PDFs)}} & \textbf{\$29.79} \\
\hline
\hline
\makecell{\textbf{Total Pre-processing} \\ \textbf{(Extrapolated)}} & \textbf{\$1,408.60} \\
\hline
\end{tabular}
\caption{A breakdown of compute for the GovScape pre-processing pipeline. All compute is reported using AWS \texttt{g4dn.4xlarge} instances (each with 16 Intel Cascade Lake vCPUs \& 1 Nvidia T4 GPU) and, with keyword index building, m5.4xlarge instances (each with 16 Intel Xeon Platinum 8175 vCPUs) in the \texttt{us-east-2} region. 
We list index building separately because each index can only be built by one server at a time. For other steps, we extrapolate our cost using a previous run of 4,736,080 PDFs, which we had configured for benchmarking. We note that this keyword index building cost reflects an earlier build using SQLite FTS5; given our results in Table \ref{tab:keyword_benchmark}, building our current Lucene keyword index is cheaper. 
}\label{tab:pipelinecompute}
\end{table}

\subsection{PDF Pre-processing: Compute \& Timing}\label{sec:compute}

In Table \ref{tab:pipelinecompute}, we break down the computing specifications, runtime, and cost for our pre-processing pipeline. Using AWS \texttt{g4dn.4xlarge} instances (each with 16 Intel Cascade Lake vCPUs and 1 Nvidia T4 GPU), we parallelized our pre-processing pipeline. We performed keyword text index building on \texttt{m5.4xlarge} instances. The estimated total cost of the compute required for our pre-processing pipeline was \$1,408.60.\footnote{All costs are computed using the $\$1.204$/hr on-demand cost for a \texttt{g4dn.4xlarge} instance and $\$0.768$/hr on-demand cost for a \texttt{m5.4xlarge} instance.} We report \$1,500 to account for our extrapolation.\footnote{This cost did not factor in associated AWS S3 fees, which are specific to our cloud computing setup. We incurred of order \$1,000 in fees writing to our S3 bucket. The network transfer fees between S3 and the EC2 instances were $\$0$ because our compute and storage were co-located, and AWS does not charge for data transfer within a region.}

\section{Search Affordances}\label{sec:search}

All three search affordances detailed below can be combined with filter conditions against metadata.

\subsection{Semantic Text Search}\label{sec:nlsearch}

For our semantic search functionality, we leverage the \texttt{BAAI/bge-base-en-v1.5} embedding model from our pre-processing pipeline. We vectorize an end-user's natural language query such as: ``economic data pertaining to geologic surveys'' by embedding the query on the back-end. We then compare this embedding to the pre-computed text embeddings for all other PDF pages and retrieve the nearest neighbors using Faiss. Here, all PDF pages are ranked according to similarity (unlike exact text search, which filters results). Again, we stress that this semantic search functionality is distinct from exact text search and enables end-users to define more expressive and flexible queries. An example of semantic text search can be found in Figure \ref{fig:semantic_text_example}.

\subsection{Visual Search}\label{sec:visualsearch}

For our visual search functionality, we leverage the \texttt{openai/clip-vit-base-patch32}  embedding model from our pre-processing pipeline \citep{CLIP_paper}. Trained using contrastive language-image pre-training (CLIP), this model jointly embeds image and text data. We can embed any natural language query over the visual features of the PDFs and perform a nearest-neighbors search in order to retrieve relevant results. For example, an end-user can formulate queries such as: ``redacted documents'' or ``aerial photographs.'' Here, we perform the analogous procedure as for semantic text search: on the back-end, each query is embedded using this CLIP model; we then compare this embedding to the pre-computed CLIP embeddings for all other PDF page images and retrieve the nearest neighbors using Faiss. As with semantic text search, all PDF pages meeting filter conditions are ranked according to similarity. Figure \ref{fig:visual_example} shows an example of visual search.

\subsection{Exact Text Search}\label{sec:exacttextsearch}
We use Lucene to perform exact text search. This extension creates an on-disk inverted document index and supports both phrase and keyword search. Due to its on-disk nature, this search is slower than the in-memory FAISS indices that support semantic search.

\section{System Architecture}

\begin{figure}[t]
    \centering
    \includegraphics[width=0.95\linewidth]{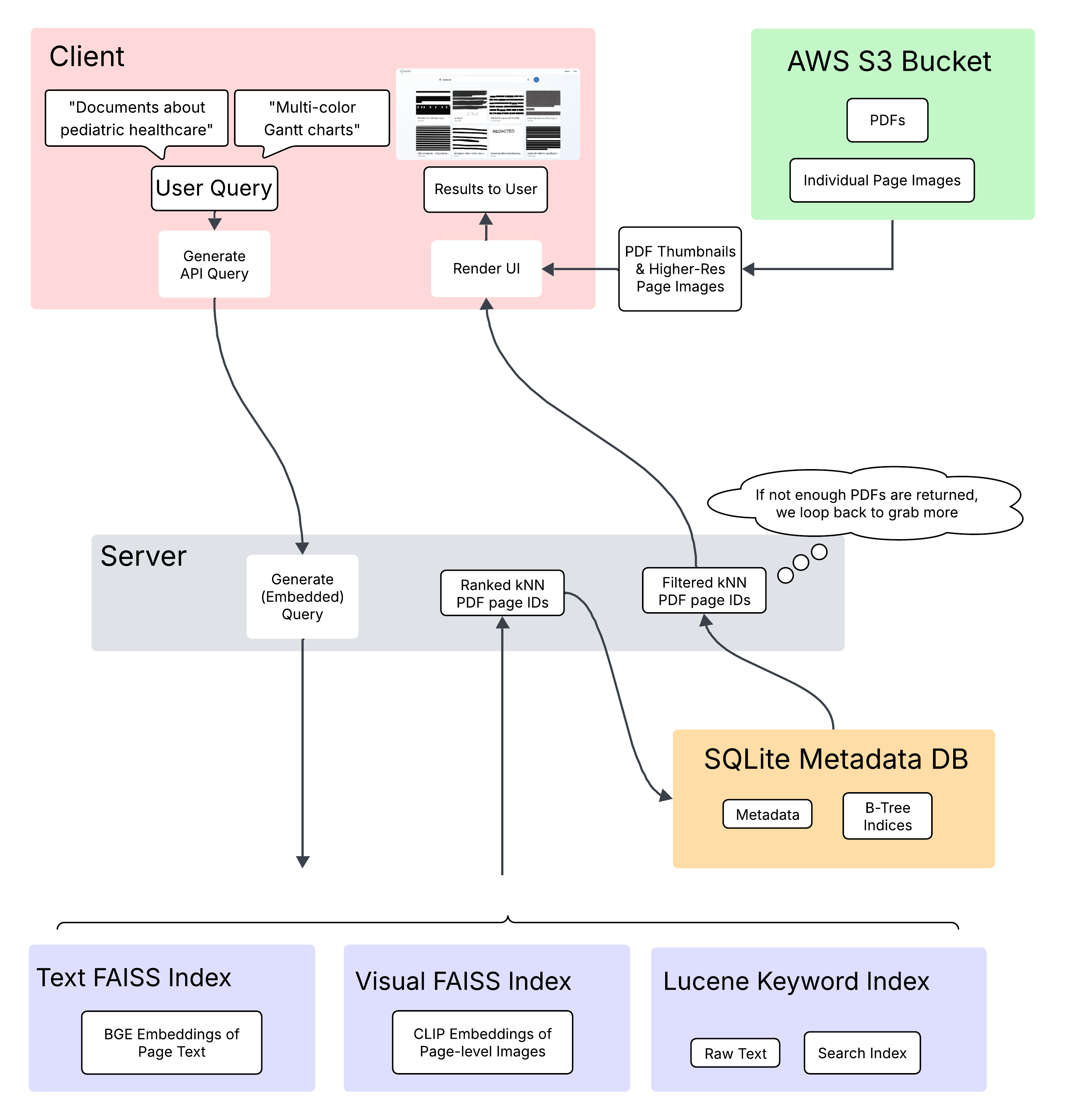}
    \caption{An overview of the GovScape architecture, showing how the constituent parts of the system interact.}
    \label{fig:architecture}
\end{figure}

The full architecture of our GovScape implementation can be found in Figure \ref{fig:architecture}. The client (top-left) sends a query to the server (middle). If the query is a semantic text search or visual search, the query is then embedded server-side and sent to Faiss (bottom-left), which returns the top $k$ nearest-neighbor results using the logic described in Section \ref{sec:nlsearch} or \ref{sec:visualsearch}, respectively. For exact text search, the requests are handled by our Lucene keyword index (bottom), as described in Section \ref{sec:exacttextsearch}. These results are then sent to the SQLite Metadata DB 
bucket (bottom-right) to filter PDFs based on filter conditions. The IDs of these filtered PDFs are sent to the client. PDF thumbnails and higher-resolution page images are retrieved from the AWS S3 bucket, and the search results are rendered for the end-user. In this section, we further detail these constituent components. Our open-source code for GovScape is available at: {\color{blue}{\url{https://github.com/govscape/govscape/}}}.

\subsection{Back-end}
The back-end of GovScape uses a Flask API Server architecture, with serving managed by Gunicorn and HTTP routing managed by nginx. Queries are handled via a six-step process:
\begin{enumerate}
\itemsep0em 

    \item \textbf{HTTP Handling:} The nginx server receives the query from the front-end and routes it to Gunicorn. It also handles functionality like encryption and rate-limiting.
    \item \textbf{Load Balancing:} The Gunicorn server sends the request to one of its threads that are running the application code, handling load balancing across CPU cores.
    \item \textbf{Embedding:} If the query type is semantic text or visual, the handler thread embeds the query using the corresponding embedding model.
    \item \textbf{Index Lookup:} Using the (potentially embedded) query, the index corresponding to the search type is queried to retrieve the top-$k$ most relevant results. \label{list:search_index}
    \item \textbf{Filtering:} These results are filtered against the metadata stored in a SQLite database. If there are too few filtered results, $k$ is doubled and we repeat the previous step.
    \item \textbf{Packaging:} The results are packaged into an HTTP response and sent back to the front-end.
\end{enumerate}
During this process, the back-end server only requires access to the indices constructed over the data, allowing it to avoid accessing the archived data in the S3 bucket. Instead, it constructs S3 addresses for the images and full PDFs, enabling the client to perform the download directly from S3. This maximizes the throughput of the back-end server by distributing the network transfer across all client devices. 

\subsection{Front-End}
The front-end architecture of GovScape is conceptually informed by the modular design patterns in the Digital Collections Explorer \citep{digital_collections_explorer}, especially in its component-based structure and the abstraction of its API communication layer. However, due to the specific needs of GovScape as a large-scale, public-facing service, we built the interface using Svelte, an open-source front-end framework. Unlike frameworks that rely on a runtime Virtual DOM, Svelte compiles components at build time into optimized vanilla JavaScript, eliminating the need for runtime diffing and reconciliation. This leads to much smaller bundle sizes and improved runtime performance.

The application comprises modular, single-responsibility components.
The main application view, \texttt{+page.svelte}, serves as the primary container, orchestrating the interactions between child components. This component-based structure enables several key features for the end-user:
\begin{itemize}
\itemsep0em 
    \item \textbf{Multi-modal search interaction}: the search interface provides three distinct search modalities: \texttt{Textual} (semantic text), \texttt{Visual} (CLIP-based), and \texttt{Keyword}.
    \item \textbf{Advanced filtering}: filter conditions are managed by the \texttt{AdvancedSearch.svelte} component. This allows users to filter search results based on specific criteria.
    \item \textbf{An interactive results gallery} is rendered by the \texttt{ResultsGrid.svelte} component. This component displays a grid of document thumbnails.
    \item \textbf{Detailed document inspection} is provided by the \texttt{PDFPreview.svelte} component.
\end{itemize}

All communication with the back-end is managed by a service layer, abstracted in \texttt{src/lib/utils/fetch.js}. When a user initiates a query, the corresponding UI component calls the \texttt{apiFetch} function, which dispatches the API request to the appropriate back-end endpoint based on the selected search mode.
Upon receiving the response, the store's state -- located in \texttt{src/lib/stores} -- is updated, triggering a reactive re-render of the interface. This clear separation of concerns between UI components, state management, and API services ensures the system is both scalable and maintainable.

In Figure \ref{fig:example_searches}, we show paginated views of ranked search results for semantic text and visual searches. Once a PDF page has been selected from the search results, we also surface a detailed view of a PDF. In this view, we show each individual PDF page, along with the PDF's metadata. We also provide a button that, when clicked, downloads the PDF.

\section{Evaluation}

\subsection{Log Analysis}
In the three months between launching on November 18th, 2025, and February 18th, 2026, GovScape received 1,679 total searches across all three search types.\footnote{All of these numbers reflect full deduplication (i.e., if a user searched for the same term using the same search type on different days, it is counted only once).} In total, GovScape received 992 semantic text search queries (59\%), 271 visual search queries (16\%), and 416 keyword search queries (25\%).

Example semantic text searches include: ``budgetary data for environmental surveys,'' ``mediation prisoner litigation,'' and ``policies regarding rainwater collection.'' Example visual searches include: ``network,'' ``comics,'' and ``monument.'' 8\% of semantic text queries have at least part of the search in quotation marks (some of the queries also contain boolean logic), suggesting that some users have conflated semantic text search and keyword search. Future work includes better communicating these search types to end-users, developing a single, catch-all query type, and conducting more nuanced log analysis as more queries are performed.

\subsection{Index Benchmarking}
To explore the tradeoffs between implementations, we performed a benchmark of various open source keyword and vector indices, as shown in Tables \ref{tab:keyword_benchmark} and \ref{tab:vector_benchmark}. For keyword indices, we considered Apache Lucene, a keyword search engine that powers major systems like ElasticSearch ~\citep{ApacheLucene}; LanceDB, a recent multi-modal search engine that offers keyword search \citep{lancedb_2026}; SQLite's FTS5 keyword search extension \citep{sqlite_fts5}; and Whoosh, an open-source pure-python keyword search engine~\citep{chaput_whoosh_2016}. To test the keyword search functionality, we downloaded 500,000 pages of text from the govscape PDF corpus, inserted them into each index, then queried each index 10,000 times using real keyword queries from the search logs. Across all metrics, Apache Lucene was the most efficient index with a similar build time to SQLite but almost $40\times$ higher query throughput and $5\times$ smaller disk size. This shows the benefits of Lucene's advanced compression techniques for its inverted index.

\begin{table}[]
    \small
    \centering
    \begin{tabular}{|c|c|c|c|}
        \hline
         \textbf{Index} & \textbf{Build (s)} & \textbf{QPS} & \textbf{Size (MB)} \\
         \hline
        \textbf{Lucene}  & \textbf{610} & \textbf{418}  & \textbf{2806}\\
         \hline
        LanceDB & 930 & 134  & 28710\\
         \hline
        SQLite  & 617 & 11   & 14122\\
         \hline
        Whoosh  & T/O & T/O  & T/O\\
         \hline
    \end{tabular}
    \caption{Keyword Index Benchmark}
    \label{tab:keyword_benchmark}
\end{table}

For vector indices, we considered four different index types offered by the FAISS library for vector search~\citep{douze2024faiss}: IVF, a clustering-based index; IVFPQ, a combination of IVF with product quantization to compress the embeddings; HNSW, a widely-used graph-based index; and FLAT, the brute-force algorithm that compares all query and embedding vectors using BLAS routines. Finally, we again include LanceDB using its default vector index implementation and prefetching data into memory to provide an even footing. We tested these methods using $500,000$ embeddings from the Govscape corpus and $10,000$ query embeddings (produced from the search log). For use in Govscape, these indices must also contain a mapping from the embedding ID to the PDF page that generated it, so we also store dictionaries that hold these mappings for the FAISS methods. For LanceDB, we use its internal metadata storage mechanisms. 

Across methods, we find that the HNSW index provides the best speed and accuracy tradeoff, but suffers in memory consumption due to its lack of embedding compression. Because space is at a premium as the dataset grows and accuracy is critical for correct search results, IVFPQ was selected as Govscape's primary vector index. However, as data scales beyond memory, it will be necessary to utilize a disk-based index like LanceDB.

\begin{table}[]
    \small
    \centering
    \begin{tabular}{|c|c|c|c|c|}
        \hline
         \textbf{Index} & \textbf{Build (s)} & \textbf{QPS} & \textbf{Size (MB)}  & \textbf{Recall@10} \\
         \hline
        IVF  & 596 & 500  & 2960 & .9674\\
         \hline
        IVFPQ  & 620 & 414  & 1588 & .9829\\
         \hline
        HNSW  & 101 & \textbf{750}  & 3062 & .9998\\
         \hline
        Flat  & \textbf{1.5} & 8.0  & 2933 & \textbf{1.000}\\
         \hline
        LanceDB  & 49.8 & 125.9  & \textbf{1509} & .8689\\
         \hline
    \end{tabular}
    \caption{Vector Index Benchmark}
    \label{tab:vector_benchmark}
\end{table}
\section{Conclusion \& Future Work}

We have introduced GovScape, a public search system for 10,015,993 PDFs (70,958,487 PDF pages) from the 2020 crawl in the End of Term Web Archive. We have detailed the pre-processing pipeline and system architecture for GovScape, as well as highlighted its multimodal search affordances, including semantic text search, visual search, exact keyword search over PDF text, and  filter conditions. Our estimated compute cost for processing and multimodally embedding these 70 million PDF pages was approximately \$1,500. We have also provided two forms of evaluation: search log analysis and vector \& keyword index benchmarking. A demo video of GovScape can be found at: \url{https://youtu.be/VpyiYW0nWp4}.

We are already working to scale up GovScape to comprehensively support PDF search across all crawls in the End of Term Web Archive (including the 2024 crawl when it is publicly available). The majority of compute required is in pre-processing; our results in Table \ref{tab:pipelinecompute} suggest that this goal is achievable on a relatively modest budget. 


Many PDFs do not contain text data that can be usefully extracted using pypdfium2 (e.g., scanned documents might not contain an underlying text representation). We plan to incorporate optical character recognition (OCR) functionality 
 using olmOCR and are already working to benchmark OCR performance against our documents \citep{olmocr, olmocr2}.
 Moreover, we currently embed PDF text using an English-only model; we plan to support multilingual PDFs.

The embedding models that we have utilized are off-the-shelf models. While their performance has been strong based on our qualitative observation, we plan to quantitatively assess model performance via a range of strategies. To improve the quality of our embeddings for government documents, we plan to finetune our embedding models. We will also begin incorporating vision language models for more sophisticated forms of search \& retrieval. In the long term, we plan to expand GovScape to cover mixed file types and provide search across an even wider set of government information. 

Lastly, we plan to continue our log analysis of our deployed system in order to establish relevance-based metrics and study user intent. This analysis will inform our refinement of our search affordances and evaluation of embedding methods. In the long term, we hope to conduct in-person studies with end-users to augment our log analysis.

\section{Ethics \& Broader Impact Statement}

We have designed GovScape with broader impact in mind. We created and developed GovScape with the central goal of improving public access to information produced by the United States Federal Government. To ensure reproducibility and public availability, we have released all our code for GovScape in an open-source codebase available at: {\color{blue}{\url{https://github.com/govscape/govscape/}}}. This includes our pre-processing pipeline as well as the full GovScape search application. The full End of Term Web Archive can be found at: {\color{blue}{\url{https://eotarchive.org/data/}}}. We note that we used Copilot and Cursor while developing the GovScape codebase. For our log analysis, we received an IRB exemption from the University of Washington Human Subjects Division, ensuring ethical human subjects practices.

\section{Acknowledgments}
We thank the University of Washington's Information School and Paul G. Allen School for Computer Science \& Engineering for facilitating this collaboration and providing compute. We thank Magda Bałazińska and Moe Kayali at the University of Washington and Jake Poznanski and Kyle Lo at the Allen Institute for AI for their feedback. 

\bibliography{custom}

\newpage
\appendix

\section{Appendix 1: Related Work}
\label{sec:related-work}
\subsection{Searching Web Archives}
Monumental efforts to archive and preserve petabytes of web data -- of interest to journalists, academics, and members of the public -- have been widely successful \citep{milligan_book, milligan_averting}. However, questions of access remain. The primary mode of access remains single-URL lookup \citep{search_as_research}, as popularized by the Internet Archive's Wayback Machine. While this mode is useful if an end-user knows the specific URL(s) of archived data they are interested in studying, it is also restrictive: for example, it is not easy for an end-user to search \textit{across} webpages.

While keyword search over web archives is extremely valuable and increasingly supported -- as evidenced by the Internet Archive's text search functionality across 64 million .gov PDFs \citep{internet_archive_search} -- this method has its own intrinsic limitations. For example, keyword search requires that a string match be present within PDF text to be returned. Moreover, queries are restricted to the text that is present. Visual content, for example, can be entirely lost, requiring a relevant caption or alt-text to register during a keyword search. 

Recent efforts have provided promising new modes of access to web archives. For example, the Harvard Library Innovation Lab's open-source WARC-GPT project enables retrieval-augmented generation (RAG)-style search over web archive files, ``allow[ing] for creating custom chatbots that use a set of web archive files as their knowledge base, letting users explore collections through conversation'' \citep{warcgpt}. The Archives Unleashed toolkit is an open-source codebase that enables large-scale data extraction and analysis of web archives, similarly focusing on textual and metadata-based visualization and analysis \citep{ruest_2020, archives_unleashed_2019, archives_unleashed_2021}. The SafeDocs File Observatory application enables searching over digital documents according to low-level metadata features \citep{safedocs}. The National Library of Hungary's SolrWayBack supports reverse image search in addition to 
free text search across file types, site domain visualization, and image geo search for up to 20 terabytes of WARC data \citep{solrwayback}.  With GovScape, we narrow our focus to PDFs but widen the scope and expressivity of potential searches through different forms of multimodal queries.

\subsection{Government Documents \& PDFs}
Government documents are essential not only to journalists but also to researchers in fields ranging from media studies to history, economics, public policy, and law \citep{gitelman_paper_knowledge, declassification}. Projects such as the FOIArchive \citep{foiarchive} and the Data Liberation Project \citep{data_liberation} demonstrate the research possibilities enabled by large-scale access to government documents \citep{souza2016usingartificialintelligenceidentify, connelly2021}.
Federal .gov websites, and the documents contained within them, offer unique opportunities to study a vast range of questions of importance to the humanities, social sciences, and journalism. Questions range from performing large-scale analysis of specific federal agencies, to studying the aesthetic choices embedded within PowerPoint design \citep{amazingmilitaryinfographics}, to excavating the histories of specific born-digital files \citep{powell, owens_estess}, to understanding the role of the federal government's web presence in disseminating information, to accessing information that has otherwise been modified or outright deleted. Significantly, the PDF is central to these inquiries. In \textit{Paper Knowledge: Toward a Media History of Documents},  Lisa Gitelman devotes an entire chapter to the PDF, arguing for the importance of the file type to the history of documents \citep{gitelman_paper_knowledge}.

\subsection{End of Term Web Archive}

Produced by crawling .gov and .mil domain sites at the end of each presidential term, the End of Term Web Archive is a unique archive of the federal government's web presence in the digital age \citep{phillips_2023_JCDL}. The full End of Term Web Archive crawls for 2008, 2012, 2016, and 2020 are available through the Amazon AWS Open Data Sponsorship Program \citep{EOT_S3}. PDFs comprise a substantive fraction of the archive. In the 2008 End of Term crawl, for example, \texttt{application/pdf} is the fourth-most common MIME type, behind only \texttt{text/html}, \texttt{image/jpeg}, and \texttt{image/gif} \citep{phillips_murray}. The PDFs within the End of Term Web Archive are publicly available and are also searchable within the Internet Archive via basic text search \citep{internet_archive_search}. GovScape builds on this work to provide multimodal search functionality across 10 million of these PDFs from the 2020 End of Term crawl -- to our knowledge, all renderable PDFs 50 pages or under -- enabling more expressive, comprehensive, and flexible searches. 

\subsection{Multimodal Search in Cultural Heritage}

As described by Smits \& Wevers~\citep{smits_wevers}, the development of models such as CLIP has led to a recent ``multimodal turn'' in the digital humanities~\citep{CLIP_paper}. Research surrounding the application of multimodal models to digital collections has demonstrated new possibilities for discoverability, enabling expressive searches across visual collections \citep{mahowald_lee, smits_JOHD, smits_kestemont, wevers_photos}. GovScape builds on this body of work to provide multimodal search capabilities for government documents, including natural language queries over visual features.

\subsection{Enabling Semantic Search with Vector Indices}
Embedding models such as CLIP transform complex objects (e.g., text and images) into vectors of numbers~\citep{CLIP_paper}. Specifically, they do so with the promise that similar objects will be mapped to similar vectors. With this promise, a user's query like ``water planning'' will be embedded in a similar vector to text documents outlining state water management plans. Conceptually, search is then a two-step process: (1) embedding the user's query into a vector, and (2) identifying the documents in the dataset that have the most similar vectors. The second step, referred to as \emph{nearest neighbor search}, becomes challenging when there are millions or billions of documents in the dataset. At this scale, brute force comparison takes far too long to support interactive search.

To solve this problem, there is a large body of work on \emph{vector indices}, which build an index structure over the document vectors in order to support fast nearest neighbor search\citep{DBLP:journals/corr/abs-2401-08281,DBLP:conf/www/GollapudiKSKBRL23,DBLP:journals/debu/KrishnaswamyMS24}. These indices typically sacrifice perfect accuracy in order to provide millisecond latency. GovScape uses the open-source Faiss library to handle its nearest neighbor search~\citep{DBLP:journals/corr/abs-2401-08281}.

\end{document}